\documentclass[12pt]{article}
\usepackage{epsfig,amssymb,amsmath,psfrag,subfigure,rotate,color,wasysym}

\allowdisplaybreaks
\newcommand{\insertfig}[2]{\includegraphics[width=#1cm]{#2}}

\DeclareMathOperator*{\SumInt}{%
\mathchoice%
  {\ooalign{$\displaystyle\sum$\cr\hidewidth$\displaystyle\int$\hidewidth\cr}}
  {\ooalign{\raisebox{.14\height}{\scalebox{.7}{$\textstyle\sum$}}\cr\hidewidth$\textstyle\int$\hidewidth\cr}}
  {\ooalign{\raisebox{.2\height}{\scalebox{.6}{$\scriptstyle\sum$}}\cr$\scriptstyle\int$\cr}}
  {\ooalign{\raisebox{.2\height}{\scalebox{.6}{$\scriptstyle\sum$}}\cr$\scriptstyle\int$\cr}}
}

\def\XXint#1#2#3{{\setbox0=\hbox{$#1{#2#3}{\int}$ }
\vcenter{\hbox{$#2#3$ }}\kern-.6\wd0}}

\def \be  {\begin{equation}}
\def \ee  {\end{equation}}
\def \ba  {\begin{eqnarray}}
\def \ea  {\end{eqnarray}}
\def \baa {\begin{eqnarray*}}
\def \eaa {\end{eqnarray*}}
\def \lab #1 {\label{#1}}

\newcommand\re[1]{(\ref{#1})}
\def\d{\hbox{{d}\kern-.20em\hbox{l}}}

\def \matrix #1 {\left(\begin{array}{cc} #1 \end{array}\right)}

\newcommand \vev [1] {\langle{#1}\rangle}
\newcommand \VEV [1] {\left\langle{#1}\right\rangle}

\newcommand{\bit}[1]{\mbox{\boldmath$#1$}}

\newcommand{\ft}[2]{{\textstyle\frac{#1}{#2}}}
\numberwithin{equation}{section}
\begin{document}

\begin{titlepage}

\thispagestyle{empty}

\vspace*{3cm}

\centerline{\large \bf Twisting perturbed parafermions}
\vspace*{1cm}

\centerline{\sc A.V.~Belitsky}

\vspace{10mm}

\centerline{\it Department of Physics, Arizona State University}
\centerline{\it Tempe, AZ 85287-1504, USA}

\vspace{2cm}

\centerline{\bf Abstract}

\vspace{5mm}

The near-collinear expansion of scattering amplitudes in maximally supersymmetric Yang-Mills theory at strong coupling is governed by the dynamics of stings
propagating on the five sphere. The pentagon transitions in the operator product expansion which systematize the series get reformulated in terms of matrix 
elements of branch-point twist operators in the two-dimensional O(6) nonlinear sigma model. The facts that the latter is an asymptotically free field theory and 
that there exists no local realization of twist fields prevents one from explicit calculation of their scaling dimensions and operator product expansion coefficients. 
This complication is bypassed making use of the equivalence of the sigma model to the infinite-level limit of WZNW models perturbed by current-current interactions, 
such that one can use conformal symmetry and conformal perturbation theory for systematic calculations. Presently, to set up the formalism, we consider the O(3) 
sigma model which is reformulated as perturbed parafermions. 

\end{titlepage}

\setcounter{footnote} 0

\newpage




\section{Introduction}

\begin{figure}[t]
\begin{center}
\mbox{
\begin{picture}(0,130)(180,0)
\put(0,-230){\insertfig{17}{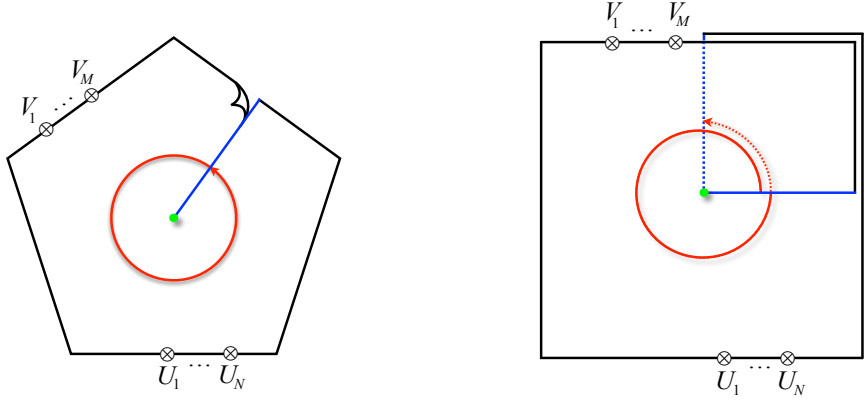}}
\end{picture}
}
\end{center}
\caption{\label{PentagonTwistsFig} The $N \to M$ particle pentagon transition (left panel) and its representation (right panel) as a twist operator insertion 
into a two-dimensional Euclidean world sheet with a branch cut running from the location of the operator insertion to infinity. The pentagon twist operator creates 
an excess angle $\ft14 (2 \pi)$.}
\end{figure}

The (regularized) S-matrix of planar maximally supersymmetric Yang-Mills theory is equivalent to the expectation value of the super Wilson loop $W_N$ living on a 
null polygonal contour with its sides determined by the on-shell external particles' momenta \cite{Alday:2007hr,Drummond:2007cf,Brandhuber:2007yx}. Making use of 
this duality, the calculation of the expectation value can be cast in a systematic framework of the operator product expansion (OPE) \cite{Basso:2013vsa,Alday:2010ku}
that allows one to determine the exact dependence on the 't Hooft coupling $\lambda = g_{\rm\scriptscriptstyle YM}^2 N_c /(4 \pi)^2$. A generic representation of 
$W_N$ for a particular tessellation reads
\begin{align}
\label{MatrixElementSuperPentagon}
W_N
=
\SumInt_{\bit{\scriptstyle U}_1,\bit{\scriptstyle U}_2, \dots, \bit{\scriptstyle U}_{N-5}}
P (0|\bit{U}_{N-5})
\dots
P (\bit{U}_2|\bit{U}_1)
{\rm e}^{- \tau_1 E_1 + i \sigma_1 P_1 + i \varphi_1 m_1}
P (\bit{U}_1|0)
\, ,
\end{align}
where the $P$'s are the pentagon transition amplitudes (see the left panel of Fig.\ \ref{PentagonTwistsFig} for a graphical representation)
\begin{align}
\label{PentagonAsMatrixElement}
P (\bit{U}|\bit{V}) =\vev{\bit{V} | \mathcal{T} | \bit{U}}
\end{align}
defined as the matrix element of an operator $\mathcal{T}$ between states of excitations propagating on the surface of the loop (or the world sheet of the evolving flux 
tube) parametrized by multidimensional vectors $\bit{U} = (\bit{u}, \bit{I})$ that cumulatively stand for particles' rapidities and their O(6) indices with respect to the 
R-symmetry group. Each of the $N-5$ intermediate states is accompanied by phases describing their propagation, encoded in their energy $E$, momentum $P$ and 
helicity $m$. From this representation, it is immediately obvious that the expansion finds its immediate application to the near-collinear limit, i.e., when $\tau_i \to \infty$, 
since one can limit oneself to the contribution with the lowest number of massive particle in each resolution identity. This is a good approximation for gluon and fermion 
excitations at any value of the coupling. However, this is not the case for scalars which develop an exponentially small mass at strong 't Hooft coupling and have to be 
resummed \cite{Basso:2014jfa,Belitsky:2015lzw,Bonini:2016knr}. These are the only propagating degrees of freedom in this regime and their dynamics is governed by 
the two-dimensional O(6) sigma model. 

The operator $\mathcal{T}$ defining the pentagon transition \re{PentagonAsMatrixElement} can be viewed as a branch-point twist operator \cite{Basso:2014jfa}. The 
latter is being known for quite some time \cite{Dixon:1986qv,Knizhnik:1987xp,Bershadsky:1987jk} and has recently found its extensive use \cite{Cardy:2007mb} in the 
computation of entanglement entropy at quantum critical points \cite{Callan:1994py,Holzhey:1994we}. However, in distinction with that application, the ``number $n$ of 
replica sheets'' for the pentagon is a fractional number (see the right panel of Fig.\ \ref{PentagonTwistsFig}), i.e., $n = \ft{5}{4}$, rather than being an integer. One finds
oneself in an immediate predicament with their use, however, as there exists no explicit realization of twist field operators in terms of elementary local fields, except for 
fermions \cite{Dixon:1986qv,Knizhnik:1987xp,Bershadsky:1987jk} and the computation of their correlation functions becomes less straightforward for asymptotically 
free theories. It is advantageous to looks at the latter as conformal field theories (CFT) perturbed by relevant operators which drive them to strong coupling in the 
infrared regime. As a consequence one can use conformal invariance of the ultraviolet theory to constrain the correlation functions and find corrections making use of 
conformal perturbation theory. In fact, the formulation of sigma models as perturbed conformal field theories was suggested a long time ago \cite{Fendley:1999gb}. In 
the present paper, we are going to start with the exploration of the O(3) sigma model which receives a dual description as perturbed model of parafermions 
\cite{Fateev:1985mm,Fateev:1990bf,Fateev:1991bv}.

\section{Field theory for parafermions}

The ${\rm O}(\mathcal{N})$ sigma model can be represented as an infinite $k$-level limit of the coset ${\rm O}_k (\mathcal{N})/{\rm O}_k(\mathcal{N}-1)$ 
Wess-Zumino-Novikov-Witten (WZNW) theory perturbed by a marginally relevant current-current operator \cite{Fateev:1990bf,Fateev:1991bv,Fendley:1999gb}. 
For the O(3) sigma model, discussed currently, the coset is a $Z_k$ parafermion theory \cite{Fateev:1990bf,Fateev:1985mm} with an action $S_{\rm CFT}$ and 
a deforming interaction determined by an addendum \cite{Fateev:1990bf,Fateev:1991bv},
\begin{align}
\delta S = - \kappa \int d^2 \bit{z} \, \bit{\Phi} (\bit{z})
\, , \qquad 
\bit{\Phi} (\bit{z})
=
\bit{\psi} (z) \bit{\bar\psi} (\bar{z}) + \bit{\psi}^\dagger (z) \bit{\bar\psi}^\dagger (\bar{z})
\, ,
\end{align}
where the measure $d^2 \bit{z} = \ft12 dz d\bar{z}$ is decomposed in terms of (anti)holomorphic variables ($\bar{z} = z_2 - i z_1$) $z = z_2 + i z_1$. The parafermion 
currents $\bit{\psi}$, $\bit{\bar\psi}$ and their charge conjugates, defined by daggers, possess the mass dimension $\Delta_k = 1 - 1/k < 1$. As a consequence, the coupling 
constant $\kappa$ has the dimension $2/k$. It is relevant and the theory flows along the renormalization group trajectory to strong coupling in the infrared. 

The original parafermion theory is completely defined by the current algebra \cite{Fateev:1990bf,Fateev:1991bv}. However, since we are interested in the analysis of 
the perturbed model, we will aim at its path integral description so as to have a local field theory with simple Feynman rules for calculation of all correlation functions. It is 
well-known that $Z_k$ CFT can be described it as a gauged model of free fermions \cite{Witten:1983ar,Bardakci:1987ee,Naculich:1989ii,Cabra:1989hg,Cabra:1996hx,Bardacki:1990wj}. 
The ${\rm O}_k (3)/{\rm O}(2) \sim {\rm SU}_k (2)/{\rm U}(1)$ coset arises from the constrained fermion fields described by the action
\begin{align}
S_{\rm CFT}
=
\frac{1}{2 \pi}
\int d^2 \bit{z} \, 
\bar\psi^a_{\alpha; m} 
\left ( 
 \delta^{ab} \delta_{n}^m {\not\!\partial}_{\alpha \beta}
+
i \delta^{ab} t_n^{I;m} {\not\!\!{G}}^I_{\alpha \beta} 
+
i \delta^{ab}  \delta_n^{m} {\not\!\!{A}}_{\alpha \beta}
+ 
i \tau_3^{ab} \delta_{n}^m {\not\!{a}}_{\alpha \beta}  
\right) \psi^{b; n}_\beta
\, .
\end{align}
Here the non-dynamical gauge fields are merely Lagrange multipliers. They are integrated out in the path integral and provide the constraints in question. Namely,
the kinetic term above possesses ${\rm U}(2k)$ internal symmetry which is encoded by the ${\rm SU} (2)$ internal symmetry indices $a = 1,2$ and $n = 1, \dots, k$
of ${\rm SU} (k)$. The first interaction term eliminates the ${\rm SU} (k)$ while the second one the ${\rm U} (1)$ currents, yielding level-$k$ ${\rm SU}_k (2)$ WZNW
theory, ${\rm SU}_k (2) \sim {\rm U}(2k)/[{\rm SU}_2 (k) \otimes {\rm U} (1)]$. Finally, the last term involving the Cartan generator of ${\rm SU} (2)$, eliminates the 
remaining ${\rm U} (1)$ and results in the ${\rm SU}_k (2)/{\rm U}(1)$ coset theory.

We can perform a transformation in the path integral to eliminate the gauge fields and thus re-express the action by means of unconstrained free fermions $\chi$.
The (anti)holomorphic components of the (non-)Abelian gauge fields $W_\mu = (G_\mu, A_\mu, a_\mu)$ can be cast in an almost ``pure gauge'' form
\begin{align}
\label{GaugeFixing}
W = i ( \bar\partial U_W ) U_W^{-1}
\, , \qquad
\bar{W} = i ( \partial \bar{U}_W ) \bar{U}_W^{-1}
\, .
\end{align}
For Abelian fields, these can be parametrized by the pseudoscalar and scalar fields $\sigma, \rho$ and $\theta, \phi$, respectively, as follows
\begin{align}
U_A = {\rm e}^{- \theta - i \sigma} \, , \qquad \bar{U}_A = {\rm e}^{\theta - i \sigma}
\, , \qquad
U_a = {\rm e}^{- \phi - i \rho} \, , \qquad \bar{U}_a = {\rm e}^{\phi - i \rho}
\, .
\end{align}
The $\sigma, \rho$ fields are pure gauges and can be ignored. Performing the variable transformation in the path integral,
\begin{align}
&
\psi_1^1 = U_A U_a U_G \chi_1^1
\, , 
&
\psi_2^1 
&
= 
\bar{U}_A \bar{U}_a \bar{U}_G
\, , 
&
\bar\psi_1^1 
&
= \bar{U}_A^{-1} \bar{U}_a^{-1} \bar{U}_G^{-1} \bar\chi_1^1
\, , 
&
\bar\psi_2^1 
&= U_A^{-1} U_a^{-1} U_G^{-1} \bar\chi_2^1
\, , \nonumber\\
&
\psi_1^2 =U_A^{-1} U_a^{-1} U_G^{-1} \chi_1^2
\, , 
&
\psi_2^2 
&
= \bar{U}_A^{-1} \bar{U}_a^{-1} \bar{U}_G^{-1} \chi_2^2
\, , 
&
\bar\psi_1^2 
&
= \bar{U}_A \bar{U}_a \bar{U}_G \bar\chi_1^2
\, , 
&
\bar\psi_2^2 
&
= U_A U_a U_G \bar\chi_2^2
\, , 
\end{align}
one can rewrite the partition function in terms of free fermion and scalar fields. We neglected here the ${\rm SU} (k)$ index to avoid cluttering formulas. The 
above transformation is accompanied by a Jacobian\footnote{It can be easily seen to arise from the path integral, schematically
$$\int D \psi D \bar\psi \, {\rm e}^{- \int d^2 z \, \bar\psi {\not D} \psi} = \frac{\det {\not\!\!D}}{\det {\not\!\partial}} 
\int D \chi D \bar\chi \, {\rm e}^{- \int d^2 z \, \bar\chi {\not \partial} \chi}.$$}. 
For the Abelian fields, it simply yields an extra scalar kinetic term, while for the non-Abelian ones, it produces the WZNW action 
\cite{Polyakov:1983tt,GamboaSaravi:1981zd} with the central charge $C_{\rm WZNW} = 2(k + 1)(k^2 - 1)/(k + 2)$. The gauge fixing procedure \re{GaugeFixing} induces 
a ghost action \cite{Karabali:1988au} with $C_{\rm ghost} = -2(k^2 + 1)$. We will not need their explicit form in what follows.

Rescaling the scalar fields $\phi \to \phi/\sqrt{k}$ and $\theta \to \theta/\sqrt{k}$, we get the canonically normalized kinetic terms in the resulting action, which read
\begin{align}
S_{\rm CFT}
=
S_{\rm free} + S_{\rm ghost} + S_{\rm WZNW}
\, ,
\end{align}
with
\begin{align}
S_{\rm free}
=
\frac{1}{\pi}
\int d^2 \bit{z}
\left( 
\partial \theta \bar\partial \theta
+
\partial \phi \bar\partial \phi
+
\bar\chi_1^a \partial \chi_2^a + \bar\chi_2^a \bar\partial \chi_1^a
\right)
\, .
\end{align}
The central charge associated with it equals the number of degrees of freedom $C_{\rm free} = 2 (k+1)$. Combining the central charges for all components, one 
recovers the one for parafermions
\begin{align}
\label{CentralChargeCFT}
C_0 = C_{\rm free} + C_{\rm ghost} + C_{\rm WZNW} = 2 \frac{k - 1}{k + 2}
\, .
\end{align}

Finally, all operators in the parafermion theory will be built from gauge-covariant fermions \cite{Bardacki:1990wj,Cabra:1994bt} by attaching a semi-infinite 
Wilson line to them stretching from their position all the way to infinity
\begin{align}
\Psi (\bit{z}) 
&= \exp \left( - i \int_{\bit{\scriptstyle z}}^\infty d \bit{z} \cdot \bit{a} (\bit{z}) \tau_3 \right) \psi (\bit{z}) 
\, , 
\end{align}
which, when expressed in terms of free fermions, read cumulatively
\begin{align}
&
\Psi^a_1 = {\rm e}^{\sigma_a \varphi (z)/\sqrt{k}} \chi^a_1 (z)
\, , \qquad
\Psi^a_2 
= {\rm e}^{- \sigma_a \bar\varphi (\bar{z})/\sqrt{k}} \chi^a_2 (\bar{z})
\, , \\
&
\bar\Psi^a_1 
= {\rm e}^{\sigma_a \bar\varphi (\bar{z})/\sqrt{k}} \bar\chi^a_1 (\bar{z})
\, , \qquad
\Psi^a_2 
= {\rm e}^{- \sigma_a \varphi (z)/\sqrt{k}} \bar\chi^a_2 (z)
\, , \nonumber
\end{align}
with the sign factor $\sigma_a$ being $(-1)^a$ and $\phi (z, \bar{z}) = \varphi (z) + \bar{\varphi} (\bar{z})$. The parafermion currents are written in their terms as
\begin{align}
\bit{\psi} (z) 
&
= 
\frac{1}{\sqrt{k}}
\Psi^2_1 (z) \bar\Psi^1_2 (z) 
\, , \qquad
\bit{\bar\psi} (\bar{z})
= 
\frac{1}{\sqrt{k}}
\Psi_2^2 (\bar{z}) \bar\Psi_1^1 (\bar{z}) 
\, , \\
\bit{\psi}^\dagger (z) 
&
= 
\frac{1}{\sqrt{k}}
\Psi^1_1 (z) \bar\Psi^2_2 (z) 
\, , \qquad
\bit{\bar\psi}^\dagger (\bar{z})
= 
\frac{1}{\sqrt{k}}
\Psi_2^1 (\bar{z}) \bar\Psi_1^2 (\bar{z}) 
\, , \nonumber
\end{align}
where the ${\rm SU} (k)$ labels are not shown and are implied to be contracted pairwise. This form immediately demonstrates the advantage of the fermion representation. 
First, one can immediately reproduce parafermion operator product expansion  \cite{Fateev:1991bv} by merely applying the Wick theorem to the above free-field 
representation as reviewed in Appendix \ref{ParAlgebraAppendix}. Second, in spite of the fact that constrained fermions do not endow the CFT model with an entirely free-field 
representation, as there is a leftover WZNW interacting sector, the latter will factorize from all correlations functions studied in this work. We should point out though
that there exists a true free-field representation for the theory of parafermions as a scalar theory with a background charge \cite{Gerasimov:1989mz}, however, we will 
not use it in the current work.

\section{Matching couplings}

Before we proceed with the derivation of the anomalous dimensions of the twist field operators as a function of the perturbing coupling, we have to find the relation between 
the coupling constant of the O(3) sigma model and the strength of the perturbing term in the parafermionic model, we will calculate the beta function in the latter theory 
by means of conformal perturbation theory. Employing invariance of the partition function under the renormalization group transformation,
\begin{align}
\int [DX]_{|\bit{\scriptstyle z}| < a} {\rm e}^{- S_{\rm CFT} - \delta S}
=
\int [DX]_{|\bit{\scriptstyle z}| < L} {\rm e}^{- S_{\rm CFT} - \delta S}
\, ,
\end{align}
we can immediately deduce the renormalized coupling
\begin{align}
\kappa (L) 
=
\mathcal{Z} (\kappa_0, L) \kappa_0 
\, , \qquad
\mathcal{Z} (\kappa_0, L)
=
1
+
\kappa_0^2 \, \mathcal{Z}_2 (L)
+
\dots
\, , 
\end{align}
where $\kappa_0 = \kappa (a)$ at the ultraviolet cutoff $a$, $a < L$, which can be safely set to zero. Here the one-loop correction is determined by the integral
\begin{align}
\mathcal{Z}_2  (L)
&
= 
\frac{1}{2} \int d^2 \bit{z}_1 d^2 \bit{z}_2 
\frac{
\vev{
\bit{\psi} (0) \bit{\bar{\psi}} (0) \bit{\psi} (z_1) \bit{\bar{\psi}} (\bar{z}_1) \bit{\psi}^\dagger (z_2) \bit{\bar{\psi}}^\dagger (\bar{z}_2) 
\bit{\psi}^\dagger (\infty) \bit{\bar{\psi}}^\dagger (\infty)
}_{\rm c}
}{
\vev{
\bit{\psi} (0) \bit{\bar{\psi}} (0) \bit{\psi}^\dagger (\infty) \bit{\bar{\psi}}^\dagger (\infty) 
}
}
\, ,
\end{align}
where only the connected part of the four-point function contributes to the integrand. The latter can be determined using the free-field Feynman rules emerging from 
field-theoretical representation discussed in the previous section. Introducing the dimensionless coupling $\kappa_{\rm R} (L) = L^{2 - 2 \Delta_k} \kappa (L)$, one can 
find the beta function for perturbed parafermion model as
\begin{align}
\frac{d \, \kappa_{\rm R} (L)}{d \ln L} = \alpha_1 \kappa_{\rm R} (L) +
\alpha_2 \kappa_{\rm R}^3 (L) 
+ \dots
\, , \qquad
\alpha_1 
= \frac{2}{k}
\, , \qquad
\alpha_2
= L^{1 - 4/k} \frac{d \mathcal{Z}_2}{d L}
\, ,
\end{align}
with
\begin{align}
\label{Z3}
\mathcal{Z}_2 (L) = \frac{2}{k^2} \int_{|\bit{\scriptstyle z}| \leq L} \frac{d^2 \bit{z}_1 d^2 \bit{z}_2}{|{z}_1|^{4/k} |{z}_2|^{2-4/k} |{z}_{12}|^{2-4/k}}
= \pi^2 L^{4/k} \frac{\Gamma (1 - \ft{4}{k}) \Gamma^2 (1 + \ft{2}{k})}{\Gamma(1 + \ft{4}{k}) \Gamma^2 (1 - \ft{2}{k})}
\, .
\end{align}
The explicit result for the integral was obtained following the recipe of Refs.\ \cite{Dotsenko:1994sy,Dotsenko:1988iq} (see also \cite{Ludwig:1987rk}).
It is analogous to the $\varepsilon$-expansion \cite{Wilson:1971dc}, however, instead of the analytical continuation of the space dimension \cite{'tHooft:1972fi}, one 
continues the central charge, or which is the same, the operator dimensions, so it is more alike the so-called analytical regularization \cite{Speer:68}.

For the case at hand, the regularization parameter is set by $\varepsilon = 1/k$. Rescaling the integration variables $z_2 = z_1 z$ ($\bar{z}_2 = \bar{z}_1 \bar{z}$), 
calculating the $\bit{z}_1$ integral with a cutoff and using the result provided in Eq.\ \re{J1integral} in Appendix \ref{AppendixIntegrals} for the remaining $\bit{z}$ integral, 
one obtains the right-hand side in Eq.\ \re{Z3}. The large-$k$ dependence of the $\alpha_\ell$ coefficients is in agreement with the result for the beta function of a perturbed 
CFT invariant under level-$k$ Kac-Moody symmetry broken by current-current interactions \cite{Destri:1988vb,Bondi:1988fp,Kutasov:1989dt,Ludwig:2002fu}. 

To find a relation between the O(3) sigma model coupling and the one of the parafermionic perturbation, we find the renormalization group trajectory for the latter,
which reads to leading order accuracy $(m L)^{2/k} = {\rm const}\times \pi \kappa_{\rm R} (L)$ and match it in the $k \to \infty$ limit to the scale $\Lambda$ 
of dimensional transmutation in the O(3) model,
\begin{align}
\Lambda = L g^{2 \beta_1/\beta_0^2} {\rm e}^{1/(\beta_0 g^2)} (1 + O(g^2))
\, , 
\end{align}
quoted here to two-loop order \cite{Polyakov:1975rr,Brezin:1976qa}, 
\begin{align}
\label{O3beta}
\beta (g) = \frac{d g^2(L)}{d \ln L} = - \beta_0 g^4 - \beta_1 g^6 + \dots 
\, ,
\qquad
\beta_0 = - \frac{1}{2 \pi}
\, , \qquad
\beta_1 = - \frac{1}{(2 \pi)^2}
\, .
\end{align}
The relation between the mass $m$ and $\Lambda$ in the MS scheme $m = \frac{8}{{\rm e}} \Lambda$ was found in Ref.\ \cite{Hasenfratz:1990zz}, however, the 
relative constant will be inconsequential for the matching to the order we are interested in. This results in the relation 
\begin{align}
\label{LambdaGrelation}
\kappa_{\rm R} = \frac{1}{\pi} \left( 1 - \frac{4 \pi}{k g^2} \right)
\, .
\end{align}
Notice that we would obtain the same result have we used the so-called mass-coupling relation, that was found by Fateev and Zamolodchikov for the case at hand
\cite{Fateev:1991bv}
\begin{align}
\label{MassCouplingRelation}
m = {\rm c}_k^{k/2} (\pi \kappa)^{k/2}
\, , \qquad
{\rm c}_k
=
(4 k)^{2/k} \frac{\Gamma (1 + \ft{1}{k})}{\Gamma (1 - \ft{1}{k})}
\, ,
\end{align}
identifying ${\rm const} = {\rm c}_k$ such that $\lim_{k \to \infty} \alpha_k = 1$.

\section{Perturbative instantons in running coupling}

Up to the order $\ell < k - 1$, there are only even powers of the coupling constant $\kappa$ showing up in the beta function for the perturbed model of parafermions.
As was noticed in the seminal paper \cite{Fateev:1991bv}, the order $\ell = k$ is peculiar since it provides the simplest topologically nontrivial contribution
for the vacuum energy. Here we will consider its manifestation in the renormalization of the coupling. The integral accompanying the relevant power of
the coupling $\kappa_0^{k-2}$ in the renormalization constant $\mathcal{Z} (\kappa_0, L)$ is 
\begin{align}
\label{Iintegral}
\mathcal{Z}_{k-2} (L)
=
\frac{1}{(k-1)!} \int_{|\bit{\scriptstyle z}| \leq L} d^2 \bit{z}_1 \dots d^2 \bit{z}_{k-2} 
\frac{
\vev{
\bit{\psi} (z_1) \bit{\bar{\psi}} (\bar{z}_1) \dots \bit{\psi} (z_{k-2}) \bit{\bar{\psi}} (\bar{z}_{k-2}) \bit{\psi} (0) \bit{\bar{\psi}} (0) \bit{\psi} (\infty) \bit{\bar{\psi}} (\infty)}
}{
\vev{
\bit{\psi}^\dagger (0) \bit{\bar{\psi}}^\dagger (0) \bit{\psi} (\infty) \bit{\bar{\psi}} (\infty)
}
}
.
\end{align}
The ratio of the correlation functions can be represented as a Vandermond determinant
\begin{align}
\frac{
\VEV{
\prod_{i = 1}^k \bit{\psi} (z_i) \bit{\bar{\psi}} (\bar{z}_i)}
}{
\vev{
\bit{\psi}^\dagger (z_{k-1}) \bit{\bar{\psi}}^\dagger (\bar{z}_{k-1}) \bit{\psi} (z_k) \bit{\bar{\psi}} (\bar{z}_k)
}
}
= (-1)^{k(k-1)/2} \frac{(k! )^2}{k^k} \frac{|{\rm det} z_i^{j-1}|^{-4/k}}{|z_{k-1,k}|^{4/k - 4}}
\, .
\end{align}
Because of the equivalence, $\bit{\psi}^\dagger_{\ell} = \bit{\psi}_{k - \ell}$, one can easily calculate the symmetry factor emerging from
fermions $\vev{ \prod_{i = 1}^k \left( \chi_1^2 (z) \bar\chi_2^1 \right) (z_i) } = k!$, where for consistency the condition $z^{2 k ({\rm mod} \, k)} = 1$ was implied.
Relying again on the same regularization procedure as alluded to above, it corresponds to the introduction of a $k$-dependent regularization parameter to accommodate
the case at hand
\begin{align}
\frac{1}{k} = \frac{1}{k_0} + \varepsilon_k
\, , 
\end{align}
where $k_0 = k-1$ with $1/\varepsilon_k = k(k-1)$. Rescaling the integration variables in \re{Iintegral} and evaluating the integral over the overall scale with an infrared 
cutoff $L$ yields a factor $L^{2 \varepsilon_k (k-2) (\Delta_k - 1)}/(2 \varepsilon_k (k-2) (\Delta_k - 1) )$ such that one can set $\varepsilon_k = 0$ in the accompanying 
$(k-3)$-fold integral over the remaining coordinates. The last step allows one to immediately use generic Dotsenko-Fateev integrals \cite{Dotsenko:1984nm,Dotsenko:1988}, 
which were demonstrated to reduce to products of Selberg integrals \cite{Selberg:2007}, 
\begin{align}
\mathcal{Z}_{k-2} (L)
=
\pi^{k - 2} (k-2)_3
\left[ \frac{k-1}{k} \frac{\Gamma (1 + \frac{1}{k-1})}{\Gamma (1 - \frac{1}{k-1})} \right]^{k-1}
\frac{L^{2 \varepsilon_k (k-2) (\Delta_k - 1)}}{\varepsilon_k (k-2) (\Delta_k - 2) }
\, ,
\end{align}
with the conventional notation for the Pochhammer symbol $(k-2)_3$. Performing naive subtraction of infinities, one finds the following correction to the beta-function 
\re{O3beta}
\begin{align}
\beta_{\rm inst} (g) = {\rm c}_{\rm inst} \frac{{\rm e}^{- 4 \pi/g^2}}{g^4}
\, ,
\qquad
{\rm c}_{\rm inst} = 2^{7} \pi^3 {\rm e}^{-1 - 2 \gamma_{\rm E}}
\, .
\end{align}
The dependence of the beta function on the coupling constant is consistent with explicit study of instantons in the renormalization of the coupling of the O(3) sigma model 
\cite{Evans:1982hx}.

\section{Twist operators and effective central charge}

\begin{figure}[t]
\begin{center}
\mbox{
\begin{picture}(0,150)(250,0)
\put(0,-135){\insertfig{13}{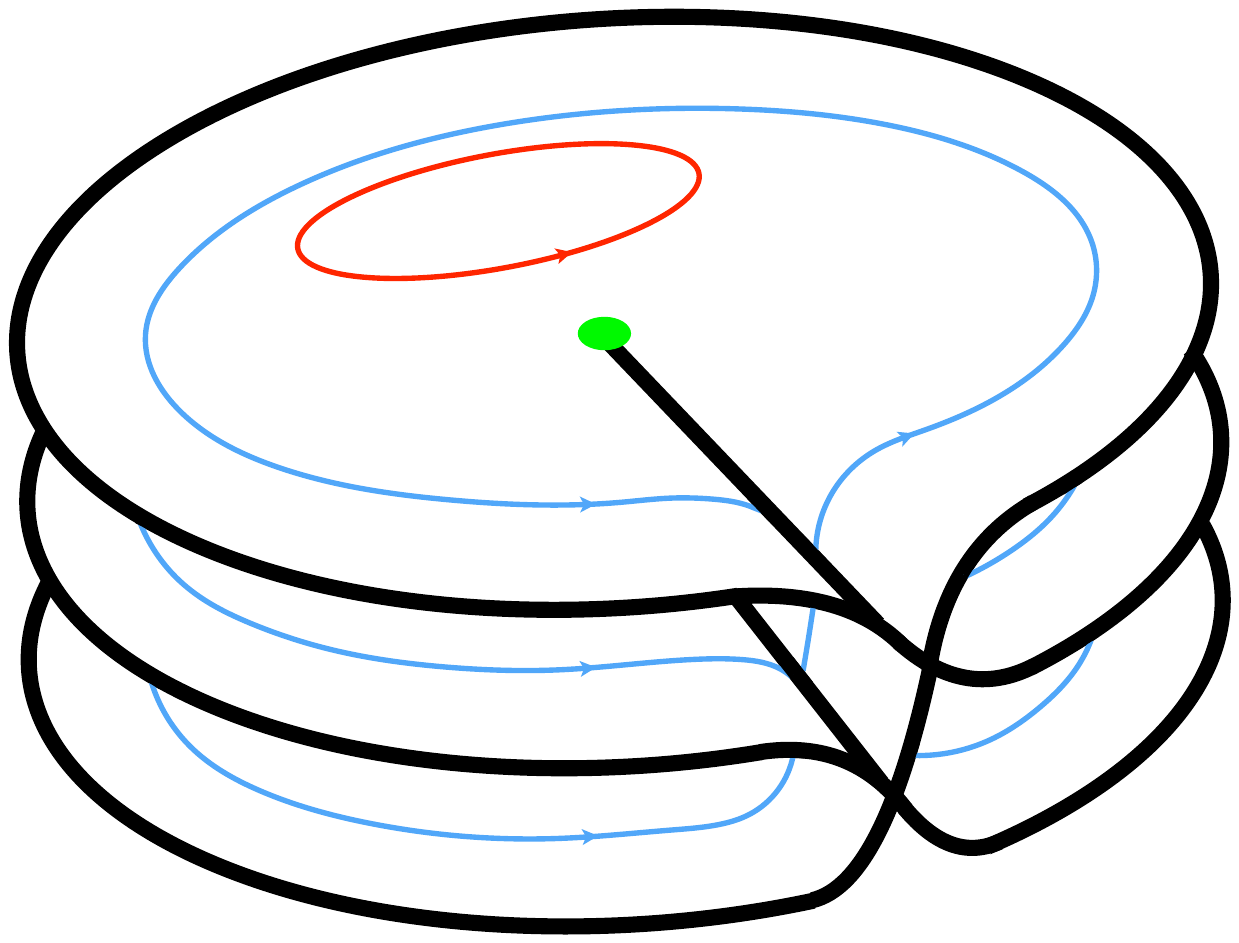}}
\put(220,-180){\insertfig{16}{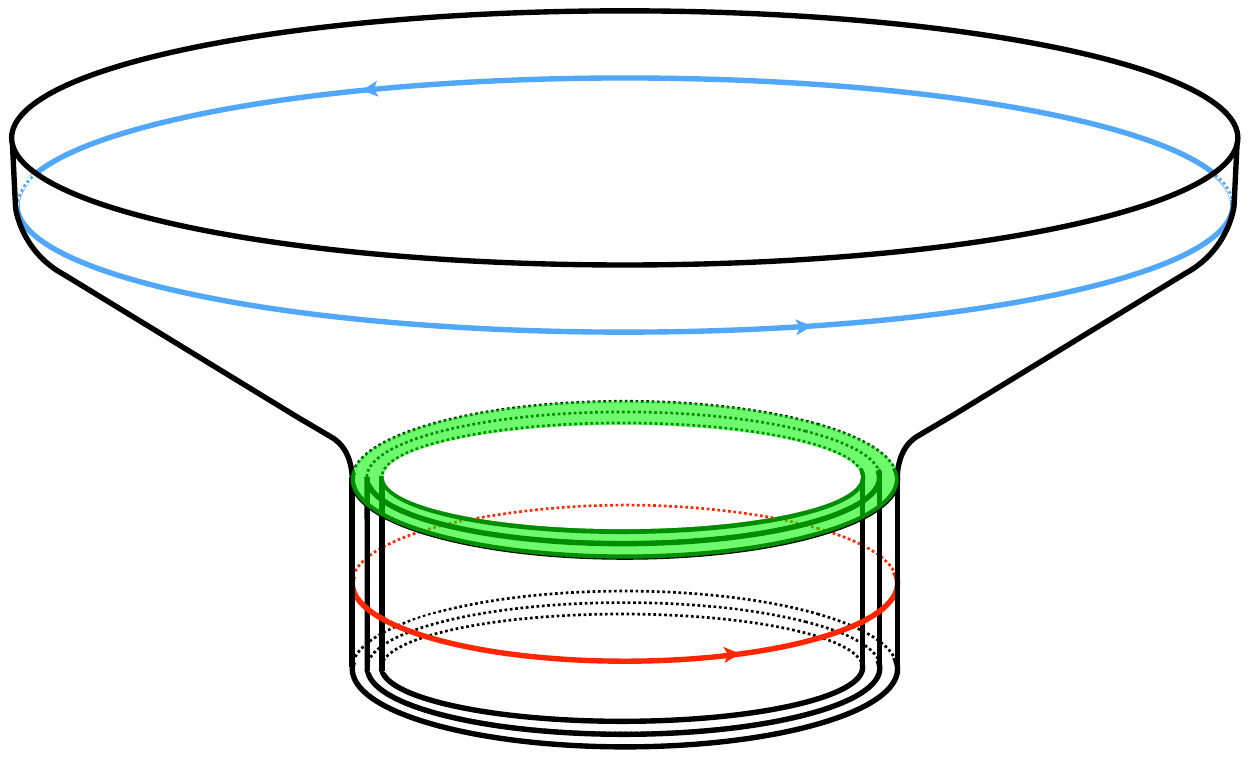}}
\end{picture}
}
\end{center}
\caption{ \label{RiemannPic} Mapping the plane replicas into the cylinder under the conformal transformation $z = {\rm e}^w$. The insertion of the twist 
operator merges $n$ copies of the cylinder of radius $R = 2 \pi$ into one of radius $2 \pi n$. In this picture for simplicity of the representation $n=3$.}
\end{figure}

As we explained in the introduction, the twist operator generating pentagon transitions create an excess angle $\ft14 (2 \pi)$, yielding a factional number $n = \ft54$ 
of two-dimensional theory replicas. Since the dependence on $n$ is anticipated (and will be found) to be an rational function, we can work with integer $n$'s
and then analytically continue to the value of interest. Thus, we consider an $n$-fold replication of the theory and introduce ramification points defined by the 
condition that when a field is brought around them it passes to the lower sheet of the Riemann surface, as shown in Fig.\ \ref{RiemannPic} for the particular case of
$n=3$. The branching point corresponds to the insertion of the twist operator $\mathcal{T}$.

To find the scaling dimension of the operator $\mathcal{T}$ in the original parafermion CFT, it proves convenient to use the operator-state correspondence and
map the complex plane into a cylinder. Conformal transformation to the cylinder $z = \exp \left( 2 \pi w/R \right)$ of radius $R$ with the periodic coordinate $w 
\sim w + i R$, yields the expectation value for the energy momentum tensor in terms of the conformal anomaly $\vev{T_{\rm cyl}} = - \frac{C}{24} (2 \pi/R)^2$ and 
thus gives the Casimir energy $E = - \frac{C}{12} (2 \pi/R)$. The dimension of the operator is given by the difference of vacuum energies $\Delta_{\mathcal{T}} = - 
R/(2 \pi)[E_{-\infty} - E_{\infty}]$. Since the initial space is $n$ copies of the cylinder the energy is $E_{-\infty} =  - n \frac{C}{12} (2 \pi/R)$, while after its insertion 
it merges the $n$ copies into a single cylinder with circumference $n R$, so that $E_{\infty} =  - \frac{1}{n} \frac{C}{12} (2 \pi/R)$, such that
\begin{align}
\Delta_{\mathcal{T}} = \frac{C}{12} \left( n - \frac{1}{n} \right)
\, .
\end{align}
Generally, the renormalization group flows of the central charge and the scaling dimension $\Delta_{\mathcal{T}}$ in the perturbed theory are different 
\cite{CastroAlvaredo:2011du} so that $\Delta_{\mathcal{T}} (\kappa) \neq C (\kappa) ( n - 1/n )/12$ as a function of the strength $\kappa$. However, the 
first term in $1/k$-series is expected to be the same since it arises from the leading deviation from the CFT behavior.

Let us calculate the effective central charge $C (\kappa)$, using finite volume effects \cite{Bloete:1986qm,Affleck:1986bv}. Its perturbative expansion reads
\begin{align}
C (\kappa) = C_0 + \sum_{\ell = 1} C_{2\ell} \kappa^{2 \ell}
\, ,
\end{align}
where the central charge of the parafermions is given in Eq.\ \re{CentralChargeCFT} and the expansion coefficients in conformal perturbation theory are 
\cite{Saleur:1987itz,Zamolodchikov:1989cf}
\begin{align}
C_{\ell}
=
24 \pi
\int_{z_1 = 1} d^2 \bit{z}_2 \dots d^2 \bit{z}_\ell 
\frac{
\vev{
\bit{\Phi}(z_1) \bit{\Phi} (\bit{z}_2) \dots \bit{\Phi }(\bit{z}_{\ell})
}
}{
|{z}_1|^{2 (1 - \Delta_k)}  \dots |{z}_{\ell}|^{2 (1 - \Delta_k)}
}
\, ,
\end{align}
with the integrand defined by the correlation function is calculated on the infinite plane and the Jacobian stemming from the map from the cylinder. There are 
only even powers of $\kappa$ showing up in the expansion up to the order $\ell = k$. The multipoint correlation functions can be easily evaluated with the free-field 
Feynman rules, e.g., 
\begin{align}
\vev{
\bit{\Phi}(\bit{z}_1) \bit{\Phi} (\bit{z}_2)
}_{\rm c}
&
= \frac{1}{|z_{12}|^{4 - 4/k}}
\, , \\
\vev{
\bit{\Phi}(\bit{z}_1) \bit{\Phi} (\bit{z}_2) \bit{\Phi}(\bit{z}_3) \bit{\Phi} (\bit{z}_4) 
}_{\rm c}
&
=
\frac{24}{k^2} \frac{|z_{13}|^{4/k - 2} |z_{14}|^{4/k - 2} |z_{23}|^{4/k - 2} |z_{24}|^{4/k - 2}}{
|z_{12}|^{4/k}
|z_{34}|^{4/k}
}
\, ,
\end{align}
such that the first two terms in the $\kappa^2$ expansion arise from the integral 
\begin{align}
\int_{z_1 = 1} d^2 \bit{z}_2 \frac{\vev{\bit{\Phi}(\bit{z}_1) \bit{\Phi}(\bit{z}_2)}
}{
|{z}_1|^{2 (1 - \Delta_k)} |{z}_2|^{2 (1 - \Delta_k)}}
=
\frac{\Gamma^2 (1 - \ft{1}{k}) \Gamma (- 1 + \ft{2}{k})}{\Gamma^2 (\ft{1}{k}) \Gamma (2 - \ft{2}{k})}
\, ,
\end{align}
while the second one is deferred to Appendix \ref{AppendixIntegrals} due its lengthy form. Since we are after the large-$k$ limit, all we have to do for higher order 
terms is to estimate their asymptotics as $k \to \infty$. We find for even $\ell$
\begin{align}
C_{\ell > 0} = - \frac{12 \pi^\ell}{k} + O(1/k^3)
\, .
\end{align}
Summing these up, together with the $O(1/k)$ term emerging from the expansion of the central charge \re{CentralChargeCFT}, one gets
\begin{align}
C (\kappa) = 2 - \frac{6}{k} \frac{1 + (\pi \kappa)^2}{1 - (\pi \kappa)^2} + O(1/k^3)
\, .
\end{align}
After substitution of the relation between parafermionic and sigma model couplings \re{LambdaGrelation} and taking the limit, one deduces
\begin{align}
C_{\rm O(3)} (g)
=
\lim_{k \to \infty} C (\kappa) 
= 
2 - \frac{3 g^2}{2 \pi} + \dots
\, .
\end{align}
This result, obtained here by resummation of conformal perturbation series, is in agreement with the result of Ref.\ \cite{Basso:2011rc} derived from the large-$\mathcal{N}$ 
study of the O($\mathcal{N}$) nonlinear sigma model \cite{Balog:2001sr,Warringa:2006rn}. The expression of the effective central charge and, as a consequence, of the 
anomalous dimension of the twist field can be cast in the form of the expansion in terms of inverse 't Hooft constant $\lambda$ by means of its relation to the O(3) 
coupling $g$, $g^2 = 1/(2 \sqrt{\lambda})$.

\section{Conclusions}

In this paper, we used the duality of the O(3) model to the perturbed model of parafermions as a means to compute anomalous dimensions of branch-point twist 
operators. Starting with CFT in the ultraviolet, we used conformal perturbation theory to calculate its flow towards strong coupling in the infrared. In fact, to determine 
the leading order correction in the original sigma model coupling $g$, we had to resum all terms in the $\kappa$ series accompanying the leading $1/k$ asymptotics 
in the Kac-Moody level $k$. This makes the derivation of higher order terms in $g$ series quite involved. 

The consideration of this work can be generalized to any $\mathcal{N}$ in O($\mathcal{N}$) making use of the results of Refs.\ \cite{Fendley:1999gb,Gerasimov:1989mz}. 
However, it appears more fruitful to pursue the route of finding an explicit realization of branch-point twist operators in terms of elementary fields building up the Lagrangian. 
While the fermionic sector was known since the inception \cite{Dixon:1986qv,Knizhnik:1987xp,Bershadsky:1987jk}, a proposal for scalar fields was put forward only recently 
\cite{Blondeau-Fournier:2016rtu}. We will report on this analysis elsewhere.

\vspace{1cm}

We would like to thank Ivan Kostov, Valya Petkova and Arkady Tseytlin for insightful communications. A part of this work was done while the author
was visiting Universita di Bologna and he would like to thank Alfredo Bonini, Davide Fioravanti, Simone Piscaglia, Francesco Ravanini and Marco Rossi 
for warm hospitality and numerous discussions. This research was supported by the U.S. National Science Foundation under the grant PHY-1403891.

\appendix

\section{Parafermion algebra}
\label{ParAlgebraAppendix}

The free-field representation 
\begin{align}
\bit{\psi}_\ell (z) 
&
= 
\sqrt{\frac{(k-\ell)!}{k! \ell!}}
\, 
{\rm e}^{+ 2 \ell \varphi (z)/\sqrt{k}} \left( \chi_1^2 (z) \bar\chi_2^1 \right)^\ell (z)
\, , \\
\bit{\psi}^\dagger_\ell (z) 
&
= 
\sqrt{\frac{(k-\ell)!}{k! \ell!}}
\, 
{\rm e}^{- 2 \ell \varphi (z)/\sqrt{k}} \left( \chi_1^1 (z) \bar\chi_2^2  \right)^\ell (z)
\, , \nonumber\\
\bit{\bar\psi}_\ell (\bar{z})
&
= 
\sqrt{\frac{(k-\ell)!}{k! \ell!}}
{\rm e}^{-2 \ell \bar\varphi (\bar{z})/\sqrt{k}} \left( \chi_2^2 (\bar{z}) \bar\chi_1^1 \right)^\ell (\bar{z})
\, , \nonumber\\
\bit{\bar\psi}_\ell^\dagger (\bar{z})
&
= 
\sqrt{\frac{(k-\ell)!}{k! \ell!}}
{\rm e}^{+ 2 \ell \bar\varphi (\bar{z})/\sqrt{k}} \left( \chi_2^1 (\bar{z}) \bar\chi_1^2 \right)^\ell (\bar{z})
\, , \nonumber
\end{align}
allows one to easily calculate the operator algebra \cite{Fateev:1985mm}
\begin{align}
\bit{\psi}_{\ell_1} (z_1) \bit{\psi}_{\ell_2} (z_2)
&
=
C_{\ell_1, \ell_2} z_{12}^{- 2 \ell_1 \ell_2/k} \bit{\psi}_{\ell_1 + \ell_2} (z_2)  [1 + O (z_{12})]
\, , \\
\bit{\psi}_{\ell_1} (z_1) \bit{\psi}^\dagger_{\ell_2} (z_2)
&
=
C_{\ell_2, k - \ell_1}
z_{12}^{- 2 \ell_2 (k - \ell_1)/k} \bit{\psi}_{\ell_1 - \ell_2} (z_1)  [1 + O (z_{12})]
\, , \nonumber\\
\bit{\psi}_{\ell} (z_1) \bit{\psi}^\dagger_{\ell} (z_2)
&
=
z_{12}^{- 2 \ell (k - \ell)/k}  [1 + O (z_{12})]
\, , \nonumber
\end{align}
where the structure constants are
\begin{align}
C_{\ell_1, \ell_2}
=
\left[
\frac{(k - \ell_1)! (k - \ell_2)!}{k! (k - \ell_1 - \ell_2)!}
\frac{
(\ell_1 + \ell_2)!}{\ell_1! \ell_2!}
\right]^{1/2}
\, .
\end{align}

\section{A couple of integrals}
\label{AppendixIntegrals}

The two-point correlation function generates the integral
\begin{align}
\label{J1integral}
\mathcal{J}_1
= 
\int d^2 \bit{z}
|z|^{2a} |1 - z|^{2 b}
\, .
\end{align}
There are a several methods one can use to calculate integrals of this type. One is based on conventional Schwinger parametrization of 
``propagators'' building up the integrand. Namely, making use of 
\begin{align}
|z|^{2 a} = \frac{1}{\Gamma (-a)} \int_0^\infty d \alpha \alpha^{- a - 1} {\rm e}^{- \alpha |z|^2}
\, .
\end{align}
Substituting this parametrization for both factors in the integrand, decomposing it in terms of holomorphic and antiholomorphic variables,
$d^2 \bit{z} = \ft12 dz d\bar{z}$ and integrating the result with respect to, say, $\bar{z}$, one immediately gets
\begin{align}
\mathcal{J}_1 = \frac{\pi}{\Gamma (-a) \Gamma(-b)} \int_0^\infty d \alpha_1 \alpha_1^{- a - 1} \int_0^\infty d \alpha_2 \alpha_2^{- b - 1} 
\int_{- \infty}^{\infty} dz {\rm e}^{ - (1 - z) \alpha_2} \delta ( z \alpha_1 - (1 - z) \alpha_2)
\, .
\end{align}
The the positivity of the integration range of the Schwinger parameters, the argument of the delta function restricts the interval of $z$ to $0 \leq z \leq 1$.
Rescaling the $\alpha$ variables as $\alpha_1 \to \alpha_1/z$ and $\alpha_2 \to \alpha_2/(1-z)$, the integrand factorizes and one gets
\begin{align}
\mathcal{J}_1 = \frac{\pi}{\Gamma (-a) \Gamma (-b)} \int_0^1 dz z^a (1 - z)^b
\int_0^\infty d \alpha_1
\int_0^\infty d \alpha_2 \delta (\alpha_1 - \alpha_2) \alpha_1^{- a - 1} \alpha_2^{- b - 1} {\rm e}^{- \alpha_2}
\, .
\end{align}
The first integral produces the Euler beta function, while the second gives the Euler gamma function yielding
\begin{align}
\mathcal{J}_1 = \pi \frac{\Gamma (-1 - a - b)}{\Gamma (-a) \Gamma (-b)} \frac{\Gamma (1 + a) \Gamma (1 + b)}{\Gamma (2 + a + b)}
\, .
\end{align}

\begin{figure}[t]
\begin{center}
\mbox{
\begin{picture}(0,100)(260,0)
\put(0,-290){\insertfig{18}{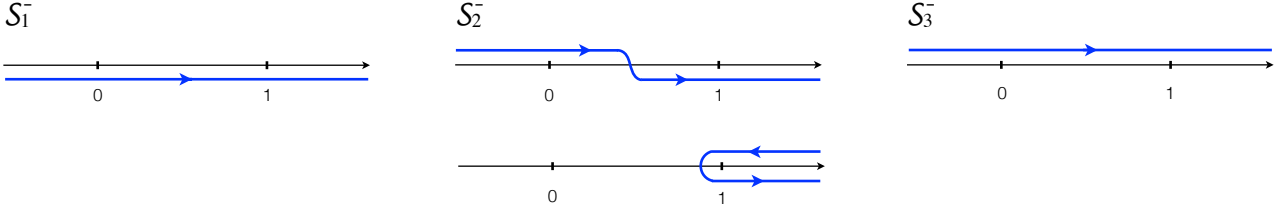}}
\end{picture}
}
\end{center}
\caption{ \label{TwoPointContours} Thee integration contours $\mathcal{S}_\alpha^-$ arising from intervals in $z_+$ variable (top row).
The deformation of the middle one picks up a discontinuity across the cut and produces nonvanishing contribution. The other two give
zero.}
\end{figure}

The other way to find the value of this integral is to perform the reduction to a product of Selberg-type integrals, see, e.g., \cite{Selberg:2007}, following the 
procedure developed by Dotsenko and Fateev \cite{Dotsenko:1984nm,Dotsenko:1988}. It starts with the analytic continuation of the integral with respect
to the second component $z_2$ of the Euclidean two-vector $\bit{z} = (z_1, z_2)$ into the complex plane with the integration path running close to the 
imaginary axis
\begin{align}
z_2 \to i z_2 (1 - i 0_+)
\, .
\end{align}
Introducing the light-cone variables $z_\pm = z_1 \pm z_2$, the $z_-$ integral carries over information on the position of branch points, or which is the same,
the deformation of the contour off the real axis,
\begin{align}
\mathcal{J}_1 = \ft{i}{2} \int_{- \infty}^{\infty} dz_+ z_+^a (1 - z_+)^b \int_{- \infty}^{\infty} dz_- [z_- + i 0_+ (z_+ - z_-)]^a [1 - z_- - i 0_+ (z_+ - z_-)]^b
\, .
\end{align}
Decomposing the entire $z_+$ axis into three intervals
\begin{align}
\mathcal{S}_1^+: \ - \infty < z_+ \leq 0
\, , \qquad
\mathcal{S}_2^+: \ 0 \leq z_+ \leq 1
\, , \qquad
\mathcal{S}_3^+: \ 1 \leq z_+ < \infty
\, ,
\end{align}
imposes constraints on the position of the branch points  and as a consequence yields the integration contours in $z_-$ shown in Fig.\ \ref{TwoPointContours}.
Only the second region gives the integration contour $\mathcal{S}_2^-$ that results in nonvanishing $z_-$ integral. Namely, by deforming the contour as shown 
in the picture below the original ones, we pick up a discontinuity such that
\begin{align}
\mathcal{J}_1 
&= 
\ft{i}{2} \int_{0}^{1} dz_+ z_+^a (1 - z_+)^b \int_{\mathcal{S}_2^-} dz_- z_-^a (1 - z_- )^b
\nonumber\\
&=
- {\rm s} (b) \int_{0}^{1} dz_+ z_+^a (1 - z_+)^b \int_1^\infty dz_- z_-^a (z_- - 1)^b
\, ,
\end{align}
where here and below ${\rm s} (\alpha) \equiv \sin(\pi \alpha)$. Making use of the well-known relation $\sin (\pi b) = - \pi/\left( \Gamma (-b) \Gamma (1 + b) \right)$, 
one verifies that the two results do coincide.

\begin{figure}[t]
\begin{center}
\mbox{
\begin{picture}(0,230)(245,0)
\put(0,-160){\insertfig{18}{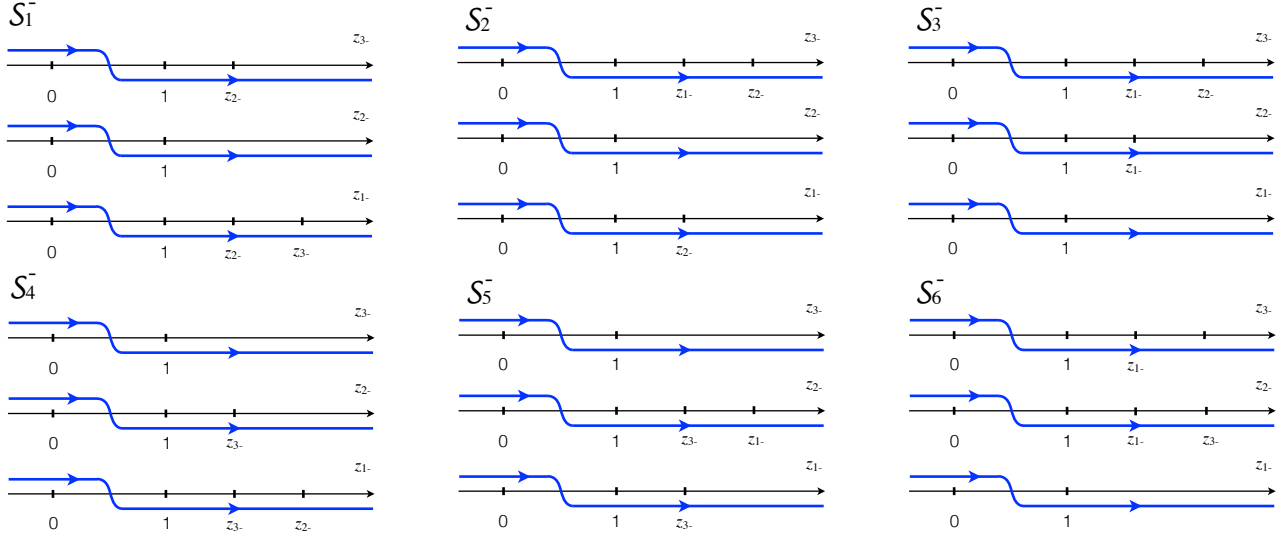}}
\end{picture}
}
\end{center}
\caption{ \label{FourPointContours} Six contours generating contributions to four-point correlation function.}
\end{figure}

Let us turn to the next term, i.e., the three-fold integral
\begin{align}
\mathcal{J}_3 
= 
\int d^2 \bit{z}_1 d^2 \bit{z}_2 d^2 \bit{z}_3 
|z_1|^{2a} |1 - z_1|^{2 b} |z_2|^{2 c} |z_3|^{2 c} |1 - z_2|^{2 d} |1 - z_3|^{2 d} |z_{12}|^{2 f} |z_{13}|^{2 f} |z_{23}|^{2g}
\, .
\end{align}
Performing the analysis identical to the above, there are six regions in $z_+$ variables that induce nonvanishing contributions
\begin{align}
&
\mathcal{S}_1^+: \ 0 \leq z_{2+} \leq 1, \quad 0 \leq z_{3+} \leq z_{2+}, \quad 0 \leq z_{1+} \leq z_{3+},
\nonumber\\
&
\mathcal{S}_2^+: \ 0 \leq z_{2+} \leq 1, \quad 0 \leq z_{3+} \leq z_{2+}, \quad z_{3+} \leq z_{1+} \leq z_{2+},
\nonumber\\
&
\mathcal{S}_3^+: \ 0 \leq z_{2+} \leq 1, \quad 0 \leq z_{3+} \leq z_{2+}, \quad z_{2+} \leq z_{1+} \leq 1,
\nonumber\\
&
\mathcal{S}_4^+: \ 0 \leq z_{2+} \leq 1, \quad z_{2+} \leq z_{3+} \leq 1,\quad 0 \leq z_{1+} \leq z_{2+},
\nonumber\\
&
\mathcal{S}_5^+: \ 0 \leq z_{2+} \leq 1, \quad z_{2+} \leq z_{3+} \leq 1, \quad z_{2+} \leq z_{1+} \leq z_{3+},
\nonumber\\
&
\mathcal{S}_6^+: \ 0 \leq z_{2+} \leq 1, \quad z_{2+} \leq z_{3+} \leq 1, \quad z_{3+} \leq z_{1+} \leq 1 ,
\end{align}
and corresponding integration contours in $z_-$ variables that are examplified in Fig.\ \ref{FourPointContours}. Deforming the latter as shown
in the middle graph in Fig.\ \ref{TwoPointContours}, we find 
\begin{align}
\mathcal{J}_3 
= 
&
- 2 {\rm s} (b) {\rm s} (d) \left[ {\rm s} (d) + {\rm s} (d + g) \right] J^-_1 J^+_1
- 2 {\rm s} (b) {\rm s} (d) \left[ {\rm s} (d + f) + {\rm s} (d + f + g) \right] J^-_1 J^+_2
\nonumber\\
&
- 2 {\rm s} (b) {\rm s} (d + f) \left[ {\rm s} (d + g) + {\rm s} (d + f + g) \right] J^-_1 J^+_3
- 2 {\rm s} (d) {\rm s} (b + f) \left[ {\rm s} (d) + {\rm s} (d + g) \right] J^-_2 J^+_1
\nonumber\\
&
- 2 {\rm s} (d) \left[ {\rm s} (b) {\rm s} (d)+ {\rm s} (b + f) {\rm s} (d + f + g) \right] J^-_2 J^+_2
- 2 {\rm s} (d) {\rm s} (b) \left[ {\rm s} (d + f) + {\rm s} (d + f + g) \right] J^-_2 J^+_3
\nonumber\\
&
- 2 {\rm s} (d) {\rm s} (b + 2 f) \left[ {\rm s} (d) + {\rm s} (d + g) \right] J^-_3 J^+_1
- 2 {\rm s} (d) {\rm s} (b + f) \left[ {\rm s} (d) + {\rm s} (d + f) \right] J^-_3 J^+_2
\nonumber\\
&\qquad\qquad\qquad\qquad\qquad\qquad\qquad\qquad\quad\,
- 2 {\rm s} (d) {\rm s} (b) \left[ {\rm s} (d) + {\rm s} (d + g) \right] J^-_3 J^+_3
\, ,
\end{align}
where there are just three Selberg-type integrals emerging from plus and minus  components each of the the two-dimensional integrals. They read
\begin{align}
J_1^+ 
&=
\int_0^1 dz_2 \int_0^{z_2} d z_3 \int_0^{z_3} dz_1 z_1^a (1-z_1)^b z_2^c (1-z_2)^d z_3^c (1-z_3)^d 
(z_2 - z_1)^f (z_3 - z_1)^f (z_2 - z_3)^g
\, , \\
J_2^+
&=
\int_0^1 dz_2 \int_0^{z_2} d z_3 \int_{z_3}^{z_2} dz_1 z_1^a (1-z_1)^b z_2^c (1-z_2)^d z_3^c (1-z_3)^d 
(z_2 - z_1)^f (z_1 - z_3)^f (z_2 - z_3)^g
\, , \\
J_3^+
&=
\int_0^1 dz_2 \int_0^{z_2} d z_3 \int_{z_2}^{1} dz_1 z_1^a (1-z_1)^b z_2^c (1-z_2)^d z_3^c (1-z_3)^d 
(z_1 - z_2)^f (z_1 - z_3)^f (z_2 - z_3)^g
\, , \\
J_1^- 
&=
\int_1^\infty dz_2 \int_1^{z_2} d z_3 \int_1^{z_3} dz_1 z_1^a (z_1-1)^b z_2^c (z_2-1)^d z_3^c (z_3-1)^d 
(z_2 - z_1)^f (z_3 - z_1)^f (z_2 - z_3)^g
\, , \\
J_2^- 
&=
\int_1^\infty dz_2 \int_1^{z_2} d z_3 \int_{z_3}^{z_2} dz_1 z_1^a (z_1-1)^b z_2^c (z_2-1)^d z_3^c (z_3-1)^d 
(z_2 - z_1)^f (z_1 - z_3)^f (z_2 - z_3)^g
\, , \\
J_3^- 
&=
\int_1^\infty dz_2 \int_1^{z_2} d z_3 \int_{z_2}^\infty dz_1 z_1^a (z_1-1)^b z_2^c (z_2-1)^d z_3^c (z_3-1)^d 
(z_1 - z_2)^f (z_1 - z_3)^f (z_2 - z_3)^g
\, .
\end{align}
For brevity, we dropped the $\pm$ subscripts from the integration variables.

Analogously, one addresses higher point integrals. The analysis is tedious but straightforward.

\end{document}